\documentclass[12pt]{article}
\usepackage{axodraw,bbold}

\catcode`@=12
\begin{document}
\thispagestyle{empty}
\begin{flushright} 
UCRHEP-T403\\ 
January 2006\
\end{flushright}
\vspace{0.5in}
\begin{center}
{\LARGE	\bf Verifiable Radiative Seesaw Mechanism\\
of Neutrino Mass and Dark Matter\\}
\vspace{1.5in}
{\bf Ernest Ma\\}
\vspace{0.2in}
{\sl Physics Department, University of California, Riverside, 
California 92521\\}
\vspace{1.5in}
\end{center}

\begin{abstract}\
A minimal extension of the Standard Model is proposed, where the observed 
left-handed neutrinos obtain naturally small Majorana masses from a one-loop 
radiative seesaw mechanism.  This model has two candidates (one bosonic and 
one fermionic) for the dark matter of the Universe.  It has a very simple 
structure and should be verifiable in forthcoming experiments at the Large 
Hadron Collider.
\end{abstract}

\newpage
\baselineskip 24pt

In the well-known canonical seesaw mechanism \cite{seesaw}, three heavy 
singlet Majorana neutrinos $N_i$ ($i=1,2,3$) are added to the Standard 
Model (SM) of elementary particles, so that
\begin{equation}
{\cal M}_\nu^{(e,\mu,\tau)} = -{\cal M}_D {\cal M}_N^{-1} {\cal M}_D^T,
\end{equation}
where ${\cal M}_D$ is the $3 \times 3$ Dirac mass matrix linking the 
observed neutrinos $\nu_\alpha$ ($\alpha=e,\mu,\tau$) to $N_i$, and 
${\cal M}_N$ is the Majorana mass matrix of $N_i$.  More generally \cite{w79}, 
${\cal M}_\nu$ comes from the unique dimension-five operator
\begin{equation}
{\cal L}_\Lambda = {f_{ij} \over \Lambda} (\nu_i \phi^0 - l_i \phi^+) 
(\nu_j \phi^0 - l_j \phi^+) + H.c.,
\end{equation}
where $(\nu_i,l_i)$ are the usual left-handed lepton doublets transforming as 
$(2,-1/2)$ under the standard electroweak $SU(2)_L \times U(1)_Y$ gauge 
group and $(\phi^+,\phi^0) \sim (2,1/2)$ is the usual Higgs doublet of the 
SM.  There are three and only three tree-level realizations \cite{m98} of 
this operator, one of which is of course the canonical seesaw mechanism. 
There are also three generic mechanisms for obtaining this operator in one 
loop \cite{m98}.  Whereas the new particles required in the three tree-level 
realizations are most likely too heavy to be observed experimentally in the 
near future, those involved in the one-loop realizations may in fact be 
light enough to be detected, in forthcoming experiments at the Large Hadron 
Collider (LHC) for example.

Consider the following minimal extension of the SM.  Under $SU(2)_L 
\times U(1)_Y \times Z_2$, the particle content is given by
\begin{eqnarray}
&& (\nu_i,l_i) \sim (2,-1/2;+), ~~~ l^c_i \sim (1,1;+), ~~~ N_i \sim 
(1,0;-), \\ && (\phi^+,\phi^0) \sim (2,1/2;+), ~~~ (\eta^+,\eta^0) \sim 
(2,1/2;-).
\end{eqnarray}
Note that the new particles, i.e. $N_i$ and the scalar doublet 
$(\eta^+,\eta^0)$, are odd under $Z_2$.
A previously proposed model \cite{m01} of neutrino mass shares the same 
particle content of this model, but the extra symmetry assumed there is 
global lepton number, which is broken explicitly but softly by the unique 
bilinear term $\mu^{2} \Phi^\dagger \eta + H.c.$ in the Higgs potential.  
Here, $Z_2$ is an exact symmetry, in analogy with the well-known $R-$parity 
of the Minimal Supersymmetric Standard Model (MSSM), hence this term is 
strictly forbidden.  As a result, $\eta^{0}$ has zero vacuum expectation 
value and there is no Dirac mass linking $\nu_i$ with $N_j$. 
Neutrinos remain massless at tree level as in the SM.

The Yukawa interactions of this model are given by
\begin{equation}
{\cal L}_Y = f_{ij} (\phi^- \nu_i + \bar \phi^0 l_i) l^c_j + h_{ij} 
(\nu_i \eta^0 - l_j \eta^+) N_j + H.c.
\end{equation}
In addition, the Majorana mass term
$${1 \over 2} M_{i} N_i N_i + H.c.$$
and the quartic scalar term
$${1 \over 2}\lambda_5 (\Phi^\dagger \eta)^2 + H.c.$$
are allowed.  Hence the one-loop radiative generation of ${\cal M}_\nu$ is 
possible, as depicted in Fig.~1.
This diagram was discussed in Ref.~\cite{m98}, but without recognizing the 
crucial role of the exact $Z_2$ symmetry being considered here.

\begin{figure}[htb]
\begin{center}
\begin{picture}(360,120)(0,0)
\ArrowLine(110,10)(150,10)
\ArrowLine(180,10)(150,10)
\ArrowLine(180,10)(210,10)
\ArrowLine(250,10)(210,10)
\ArrowLine(180,55)(150,10)
\ArrowLine(180,55)(210,10)
\ArrowLine(160,85)(180,55)
\ArrowLine(200,85)(180,55)

\Text(130,0)[]{$\nu_i$}
\Text(230,0)[]{$\nu_j$}
\Text(180,0)[]{$N_k$}
\Text(155,40)[]{$\eta^{0}$}
\Text(210,40)[]{$\eta^{0}$}
\Text(158,95)[]{$\phi^{0}$}
\Text(210,95)[]{$\phi^{0}$}

\end{picture}
\end{center}
\caption{One-loop generation of neutrino mass.}
\end{figure}
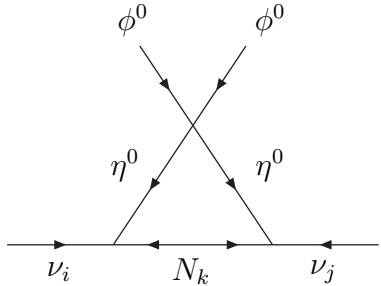

The immediate consequence of the exact $Z_2$ symmetry of this model is the 
appearance of a lightest stable particle (LSP).  This can be either bosonic, 
i.e. the lighter of the two mass eigenstates of $Re \eta^0$ and $Im \eta^{0}$, 
or fermionic, i.e. the  lightest mass eigenstate of $N_{1,2,3}$.  The latter 
possibility was first proposed in a different model \cite{knt03}, where 
neutrino masses are radiatively generated in three loops with the addition 
of two charged scalar singlets.

The Higgs potential of this model is given by
\begin{eqnarray}
V &=& m_1^{2} \Phi^\dagger \Phi + m_2^{2} \eta^\dagger \eta + {1 \over 2} 
\lambda_1 (\Phi^\dagger \Phi)^{2} + {1 \over 2} \lambda_2 
(\eta^\dagger \eta)^{2} + \lambda_3 (\Phi^\dagger \Phi)(\eta^\dagger \eta) 
\nonumber \\ &+& \lambda_4 (\Phi^\dagger \eta)(\eta^\dagger \Phi) + 
{1 \over 2} \lambda_5 [(\Phi^\dagger \eta)^{2} + H.c.],
\end{eqnarray}
where $\lambda_5$ has been chosen real without any loss of generality. 
For $m_1^{2} < 0$ and $m_2^{2} > 0$, only $\phi^{0}$ acquires a nonzero 
vacuum expectation value $v$.  The masses of the resulting physical scalar 
bosons are given by
\begin{eqnarray}
m^{2} (\sqrt 2 Re \phi^{0}) &=& 2 \lambda_1 v^{2}, \\ 
m^{2} (\eta^{\pm}) &=& m_2^{2} + \lambda_3 v^{2}, \\ 
m^{2} (\sqrt 2 Re \eta^{0}) &=& m_2^{2} + (\lambda_3 + \lambda_4 + 
\lambda_5) v^{2}, \\ 
m^{2} (\sqrt 2 Im \eta^{0}) &=& m_2^{2} + (\lambda_3 + \lambda_4 - 
\lambda_5) v^{2}.
\end{eqnarray}

The diagram of Fig. 1 is exactly calculable from the exchange of 
$Re \eta^{0}$ and $Im \eta^{0}$ and is given by
\begin{equation}
({\cal M}_\nu)_{ij} = \sum_k {h_{ik} h_{jk} M_k \over 16 \pi^{2}} \left[ 
{m_R^{2} \over m_R^{2}-M_k^{2}} \ln {m_R^{2} \over M_k^{2}} - 
{m_I^{2} \over m_I^{2}-M_k^{2}} \ln {m_I^{2} \over M_k^{2}} \right],
\end{equation}
where $m_R$ and $m_I$ are the masses of $\sqrt 2 Re \eta^{0}$ and $\sqrt 2 
Im \eta^{0}$ respectively.  If $m_R^{2} - m_I^{2} = 2 \lambda_5 v^{2}$ is 
assumed to be small compared to $m_0^{2} = (m_R^{2} + m_I^{2})/2$, then
\begin{equation}
({\cal M}_\nu)_{ij} = {\lambda_5 v^{2} \over 8 \pi^{2}} 
\sum_k {h_{ik} h_{jk} M_k \over m_0^{2} - M_k^{2}} \left[ 
1 - {M_k^{2} \over m_0^{2}-M_k^{2}} \ln {m_0^{2} \over M_k^{2}}  \right].
\end{equation}
If $M_k^{2} >> m_0^{2}$, then
\begin{equation}
({\cal M}_\nu)_{ij} = {\lambda_5 v^{2} \over 8 \pi^{2}} 
\sum_k {h_{ik} h_{jk} \over M_k} \left[ 
\ln {M_k^{2} \over m_0^{2}} - 1 \right].
\end{equation}
If $m_0^{2} >> M_k^{2}$, then
\begin{equation}
({\cal M}_\nu)_{ij} = {\lambda_5 v^{2} \over 8 \pi^{2} m_0^{2}} 
\sum_k h_{ik} h_{jk} M_k.
\end{equation}
If $m_0^{2} \simeq M_k^{2}$, then
\begin{equation}
({\cal M}_\nu)_{ij} \simeq {\lambda_5 v^{2} \over 16 \pi^{2}} 
\sum_k {h_{ik} h_{jk} \over M_k}.
\end{equation}
From the above, it is clear that the seesaw scale is reduced by roughly 
the factor $\lambda_5 /16 \pi^{2}$.  Assuming $\lambda_5 \sim h^{2} \sim 
10^{-4}$, the corresponding canonical seesaw scale of $10^{9}$ GeV (with 
$m_\nu \sim h^{2} v^{2} / M \sim 1$ eV) is then 
reduced to just 1 TeV, which is amenable to experimental verification in 
forthcoming experiments at the LHC, for example.

This radiative seesaw mechanism of neutrino mass also predicts the 
existence of dark matter, either in the form of $N_1$ (assuming $M_1 < M_2 
< M_3$) or $\sqrt 2 Re \eta^{0}$ (assuming that it is the lightest scalar 
particle odd under $Z_2$).  
In the former case, if the $\eta$ masses are 
all greater than $M_k$, there will be observable decays
\begin{equation}
\eta^{\pm} \to l^{\pm} N_{1,2,3},
\end{equation}
then
\begin{equation}
N_{2} \to l^{\pm} l^{\mp} N_1
\end{equation}
and
\begin{equation}
N_{3} \to l^{\pm} l^{\mp} N_{1,2}
\end{equation}
through $\eta^{\pm}$ exchange.  The Yukawa couplings $h_{ij}$ may then 
be extracted and compared against the neutrino mass matrix as a means 
of verifying the seesaw mechanism \cite{m01}.

In the latter case, with $\sqrt 2 Re \eta^{0}$ as a bosonic dark-matter 
candidate \cite{sdm}, the fact that $\sqrt 2 Im \eta^{0}$ must be just 
slightly heavier is a natural condition for their coannihilation in the 
early Universe \cite{olive}.  This is better than the usual supersymmetric 
scenario for dark matter, where coannihilation requires the accidental 
degeneracy of two unrelated particles.

If $M_k$ are all greater than the $\eta$ masses, there will be observable 
decays
\begin{equation}
N_{1,2,3} \to l^{\pm} \eta^{\mp},
\end{equation}
then
\begin{equation}
\eta^{\mp} \to \eta^{0} + \{W^{\mp}\},
\end{equation}
where the real or virtual $W^{\mp}$ becomes a quark or lepton pair. Again the 
Yukawa couplings $h_{ij}$ may be extracted.

The $\eta$ particles can be produced in pairs directly by the SM gauge bosons 
$W^{\pm}$, $Z$, or $\gamma$.  Their subsequent decays will produce $N_i$ 
if kinematically allowed.  In the case where $N_{1,2,3}$ are all heavier 
than the $\eta$ particles, pair production by $e^+ e^{-}$ annihilation through 
$\eta^{\pm}$ exchange appears to be the only realistic possibility.

This model is also a very suitable framework for considering lepton family 
symmetry.  It has the flexibility of having the neutrino mass matrix 
proportional to the inverse mass matrix of $N_i$ as in the canonical 
seesaw mechanism \cite{m05}, or to the mass matrix of $N_i$ itself.  For 
example, using the tetrahedral symmetry $A_4$ \cite{a4}, many recent 
ideas \cite{recent} of implementing tribimaximal mixing \cite{hps} can be 
easily incorporated.

In conclusion, with a minimal addition to the Standard Model, i.e. a second 
scalar doublet and three heavy neutral fermion singlets transforming as $-1$ 
under an exact $Z_2$ symmetry, realistic radiative neutrino masses can be 
obtained together with candidates for the dark matter of the Universe. 
This framework parallels that of the SM in family structure and the new 
particles are very likely to be observable in forthcoming experiments at 
the Large Hadron Collider, or at a future Linear Collider.

This work was supported in part by the U.~S.~Department of Energy under Grant 
No.~DE-FG03-94ER40837.

\bibliographystyle{unsrt}

\end{document}